\documentclass[superscriptaddress, onecolumn, citeautoscript, prb]{revtex4-2}
\usepackage{amsmath, amssymb, graphicx, bm}
\usepackage{xr-hyper}
\usepackage[colorlinks=true,citecolor=blue,urlcolor=blue]{hyperref}
\usepackage{xcolor, soul}
\usepackage{gensymb}
\usepackage{float}
\usepackage{bbm}
\usepackage{lipsum}
\usepackage{amsfonts,graphicx}
\usepackage{color}
\usepackage{natbib}
\usepackage{bm}
\usepackage{comment}
\usepackage{enumerate}
\usepackage{booktabs}
\usepackage{multirow}
\usepackage[version=4,arrows=pgf-filled,
textfontname=sffamily,
mathfontname=mathsf]{mhchem}

\usepackage{graphicx,subcaption,siunitx,tikz}
\usepackage[export]{adjustbox} 
\definecolor{colD}{HTML}{1F77B4}
\definecolor{col2D}{HTML}{FF7F0E}
\definecolor{col2H}{HTML}{2CA02C}
\definecolor{col2eta}{HTML}{D62728}

\usepackage{amsmath, amssymb, graphicx, bm}
\usepackage{xr-hyper}
\usepackage{hyperref}
\usepackage{xcolor, soul}
\usepackage{gensymb}
\usepackage{bbm}
\usepackage{float}
\usepackage{bbm}
\usepackage{lipsum}
\usepackage{booktabs}

\begin{document}
\setcitestyle{super}

\title{Explosive dispersal of non-motile microbes through metabolic buoyancy}

\author{Jimreeves David}
\email{jimreevesdm@ncbs.res.in}
\affiliation{National Centre for Biological Sciences, Tata Institute of Fundamental Research, Bangalore 560065, India}
\author{Shashi Thutupalli}
\email{shashi@ncbs.res.in}
\affiliation{National Centre for Biological Sciences, Tata Institute of Fundamental Research, Bangalore 560065, India}
\affiliation{International Centre for Theoretical Sciences, Tata Institute of Fundamental Research, Bangalore 560089, India}

\date{\today}

\begin{abstract}

For non-motile microorganisms, spatial expansion in quiescent fluids is presumed to be limited by diffusion. We report that microbial colonies can explosively circumvent this constraint through a self-amplifying physical process. As non-motile yeast and bacteria metabolize dense nutrients into lighter waste within their fluid environment, they generate buoyancy-driven Rayleigh-B\'enard convection, an ubiquitous fluid-dynamical phenomenon that organizes material on scales from chemical reactors to planetary atmospheres. This robust, self-generated flow fragments and disperses cellular aggregates, which seed new growth sites, enhancing total metabolic activity and further strengthening the convective flow in an autocatalytic cycle. The resulting expansion follows accelerating power-law kinetics, quantitatively captured by a physical theory linking metabolic flux to flow velocity, and produces fractal patterns through a flow-focusing instability we term Circulation-Driven Aggregation, the hydrodynamic analogue of Diffusion-Limited Aggregation. This `metabolic fireworks' mechanism establishes a canonical instance of proliferating active matter, where cellular metabolic activity self-organizes a physical transport engine--a living Rayleigh-B\'enard convection--providing a fundamental, physics-based dispersal strategy.
\end{abstract}

\maketitle

A simple biophysical reality offers non-motile organisms in quiescent fluid environments the possibility to escape the constraints of diffusive transport~\cite{Matsushita1990, BenJacob1998, Grimson1994, Rana2017, denny1993air}. In liquid surroundings, internal cellular metabolism inherently alters the external fluid density, for example, through the consumption of dense nutrients and the excretion of lighter waste~\cite{atis2019microbial, narayanasamy2025metabolically}. This sets up the conditions for a ubiquitous fluid mechanical phenomenon, a bulk solutal Rayleigh-B\'enard instability, to drive sustained convective flows~\cite{rogers2005buoyant, normand1977convective, schumacher2020colloquium, hadley1735vi, mckenzie1974convection, helffrich2001earth, bouffard2019convection, speer1995growth}. When these self-generated flows interact with frangible (microbial) cellular aggregates, the flows can become strong enough to physically tear apart and disperse the very aggregates that create them. What, then, are the resulting large-scale dispersal dynamics, and what physical principles govern the kinetics and morphogenesis of the cellular colonies?

\begin{figure}
    \centering
    \includegraphics[width=0.8\textwidth]{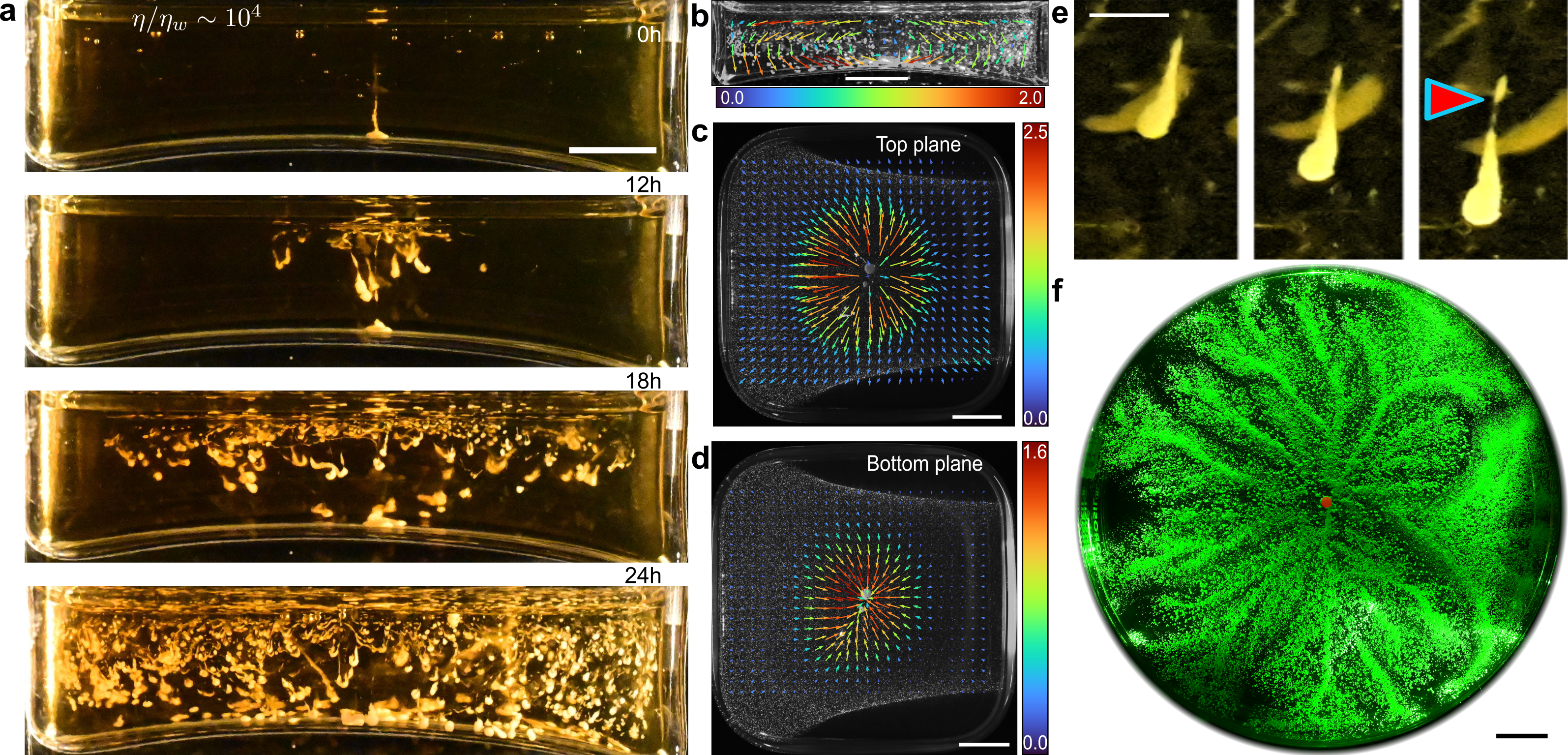}
    \caption{\textbf{The Metabolic Fireworks Phenomenon.} (\textbf{a}) Side-view time series showing the vertical stretching and subsequent dispersal of the initial yeast colony seeded at the bottom ($\eta/\eta_{w}\sim10^{4}$, movie S1). Experimental flow fields (PIV) from the (\textbf{b}) side view, (\textbf{c}) top plane (showing outward flow), and (\textbf{d}) bottom plane (showing inward flow), revealing a toroidal circulation. The colour bars show the flows speeds in units of mm/h.  (\textbf{e}) Time sequence (side view) showing flow-induced fragmentation and detachment of a colony segment (red arrow indicates a representative fragmentation event), movie S2. (\textbf{f}) The resulting 2-dimensional morphology of the colony, movie S3. H = 8~mm in \textbf{a-d}. Scale bars, 10~mm (\textbf{a, b, c, d}), 2~mm (\textbf{e}), 20~mm (\textbf{f}).}
    \label{fig:fig1}
\end{figure}

To investigate this, we inoculated a small seed colony of yeast (\textit{S. cerevisiae}) at the bottom center of a container filled with a quiescent, highly viscous nutrient medium (viscosity $\eta \approx 10^{4} \times \eta_{\text{water}}$). Time-lapse imaging from the side revealed a remarkable three-dimensional dispersal process, with the initial colony undergoing vertical stretching and fragmentation that rapidly populated the fluid volume (Fig.~\ref{fig:fig1}\textbf{a}, movie S1, fig.~\ref{figS:Controls}). This explosive dispersal is driven by self-generated convective flows. Particle Image Velocimetry revealed a toroidal circulation pattern: strong vertical plumes rise from the active biomass at the bottom boundary (Fig.~\ref{fig:fig1}\textbf{b}), turn radially outward upon reaching the top surface (Fig.~\ref{fig:fig1}\textbf{c}), and are complemented by convergent inward flow along the bottom boundary to satisfy mass conservation (Fig.~\ref{fig:fig1}\textbf{d}).

The interaction between this flow and the frangible yeast colony initiates a dispersal cycle. Close observation showed that hydrodynamic shear generated by the rising plumes is sufficiently strong to overcome cohesive forces within the colony, causing fragments, ``seeds'', to break off (Fig.~\ref{fig:fig1}\textbf{e}). These seeds are entrained within the convective loops, where they grow, undergo further shear, and generate new seedlings, that, in turn, hitchhike~\cite{luschi2003current, hays2017ocean, sale2009navigational} on the convective vortex loops~\cite{wright2008understanding, boehm2008maize, drautz2022vertical, kraitzman2024mathematical}. This autocatalytic process is a bulk 3D phenomenon, distinct from colony dynamics restricted to the air-liquid interface~\cite{atis2019microbial}, leading to rapid dispersal throughout the volume (Fig.~\ref{fig:fig1}\textbf{a}). This spread gives rise to a rich colony morphology (Fig.~\ref{fig:fig1}\textbf{f}; movies S1-S3).

\begin{figure}
    \centering
    \includegraphics[width=0.8\textwidth]{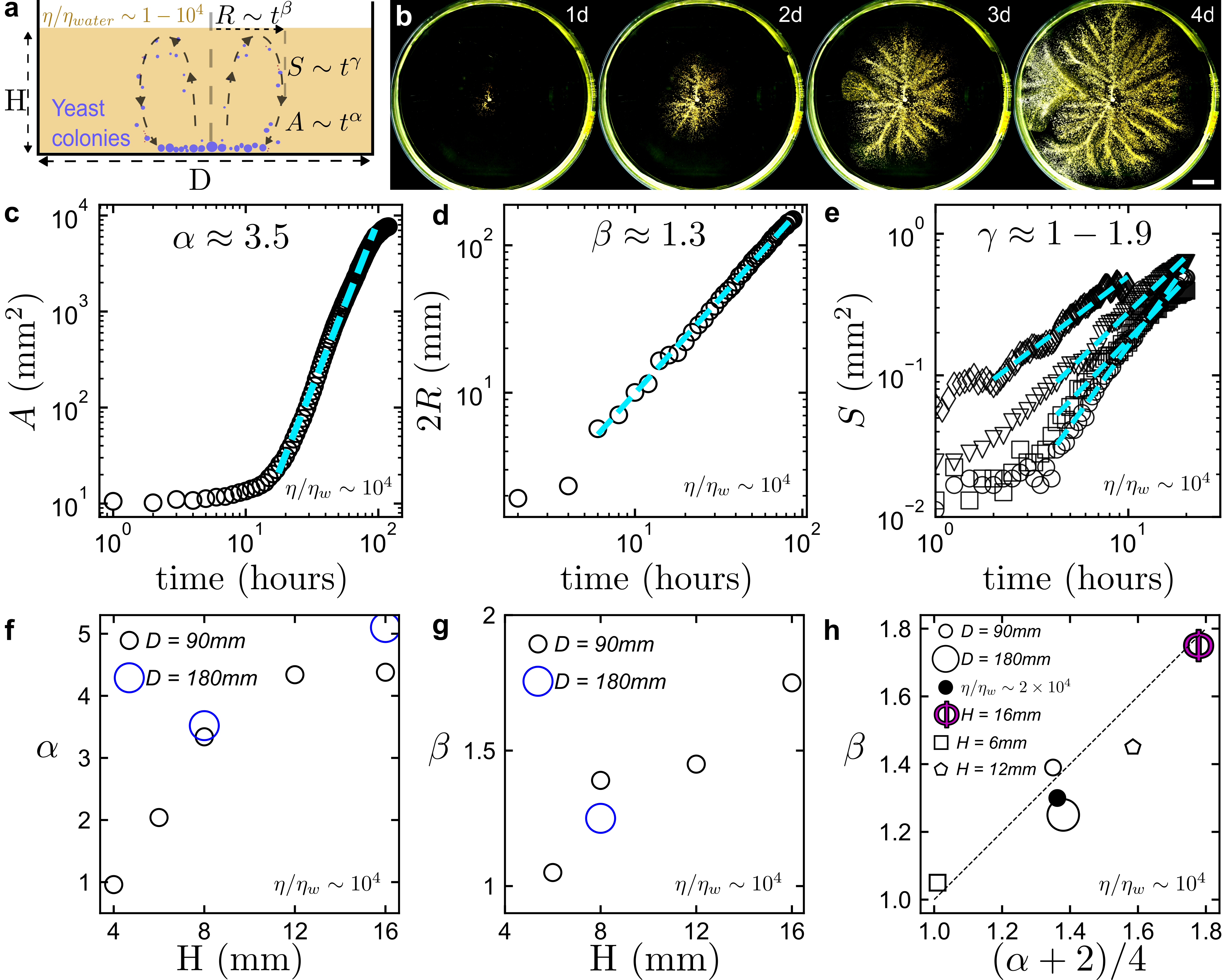}
    \caption{\textbf{Explosive expansion kinetics.} (\textbf{a}) Schematic of the experimental setup showing key parameters and observed scaling behaviors. (\textbf{b}) Time series (bottom view) showing the fractal, firework-like expansion of the colony over 4 days. The kinetics of expansion exhibit power law behaviors (H = 8~mm, $\eta/\eta_{w}\sim10^{4}$). (\textbf{c}) The total colonized area $A(t)$ grows explosively, $A(t) \propto t^{\alpha}$ ($\alpha \approx 3.5$). (\textbf{d}) The radius of the expanding front $R(t)$ shows super-linear growth, $R(t) \propto t^{\beta}$ ($\beta \approx 1.3$), indicating acceleration. (\textbf{e}) Local growth of individual seeds $S(t)$ follows a slower power law, $S(t) \propto t^{\gamma}$ ($\gamma \approx 1.0-1.9$, fig.~\ref{figS:Controls-extras}\textbf{b}). The power law exponents for (\textbf{f}) area expansion ($\alpha$) and (\textbf{g}) radial expansion ($\beta$) increase with the fluid height H, for different container diameters. (\textbf{h}) Experimental validation of the theoretical prediction $\beta = (\alpha+2)/4$ across all tested conditions (including variations in height H, diameter D, and viscosity $\eta$, movies S3-S6). Scale bars, 20~mm (\textbf{b}).}
    \label{fig:fig2}
\end{figure}

We next quantified the explosive kinetics~\cite{perez2020universal, manchein2020strong} of this process (Fig.~\ref{fig:fig2}\textbf{a}). The lateral expansion followed a firework-like pattern (Fig.~\ref{fig:fig2}\textbf{b}), with the colony area exhibiting a growth kinetics scaling as $A(t) \propto t^{\alpha}$ with $\alpha \approx 3.5$ ($\eta/\eta_{\text{water}} \sim 10^4$; Fig.~\ref{fig:fig2}\textbf{c}). The colony radius exhibited an accelerated expansion $R(t) \propto t^{\beta}$ with $\beta \approx 1.3$ (Fig.~\ref{fig:fig2}\textbf{d}), while individual seed clusters grew in a power-law manner, following $S(t) \propto t^{\gamma}$ with $\gamma \approx 1.0-1.9$ (Fig.~\ref{fig:fig2}\textbf{e}, fig.~\ref{figS:Controls-extras}). The expansion kinetics were strongly dependent on system parameters. Both area and radial growth exponents ($\alpha$ and $\beta$) increased with fluid height H for different container diameters (Fig.~\ref{fig:fig2}\textbf{F, G}, fig.~\ref{figS:Controls-extras},~\ref{figS:Height-variation}), with $\beta$ reaching $\approx 1.75$ in deeper layers.

To understand the physical principles governing these observations, we analyzed the relevant dimensionless numbers. The system operates in a distinctive viscous/convective regime characterized by low Reynolds number ($Re \ll 1$, dominated by viscous forces) and high P\'eclet number ($Pe \gg 1$, transport dominated by convection over diffusion). In this regime, a fundamental balance emerges between viscous resistance and convective transport that links the flow velocity to biomass distribution. The flow is driven by metabolically generated buoyancy with the buoyancy transport proportional to the biomass density, while viscous resistance scales with the flow velocity (Materials and Methods; Theoretical Framework in Supplementary Material).

Using the above scaling arguments, we find that the convective plume velocity $U_p$ scales with the biomass area fraction $\phi(t) = A(t)/R(t)^2$ as $U_p(t) \propto \phi(t)^{1/2}$. Since the radial expansion is driven by these plumes ($dR/dt \sim U_p$), and using the power-law growth $A(t) \sim t^\alpha$ and $R(t) \sim t^\beta$, we obtain the parameter-free scaling law:
\begin{equation}
\beta = \frac{\alpha + 2}{4}.
\end{equation}
Remarkably, this theoretical prediction, derived from first principles for the viscous/convective regime, accurately captures all experimental data across varying heights, diameters, and viscosities (Fig.~\ref{fig:fig2}\textbf{h}, movies S3-S6, fig.~\ref{figS:4cases-prt1},~\ref{figS:4-cases-extras2}).

\begin{figure}
    \centering
    \includegraphics[width=0.8\textwidth]{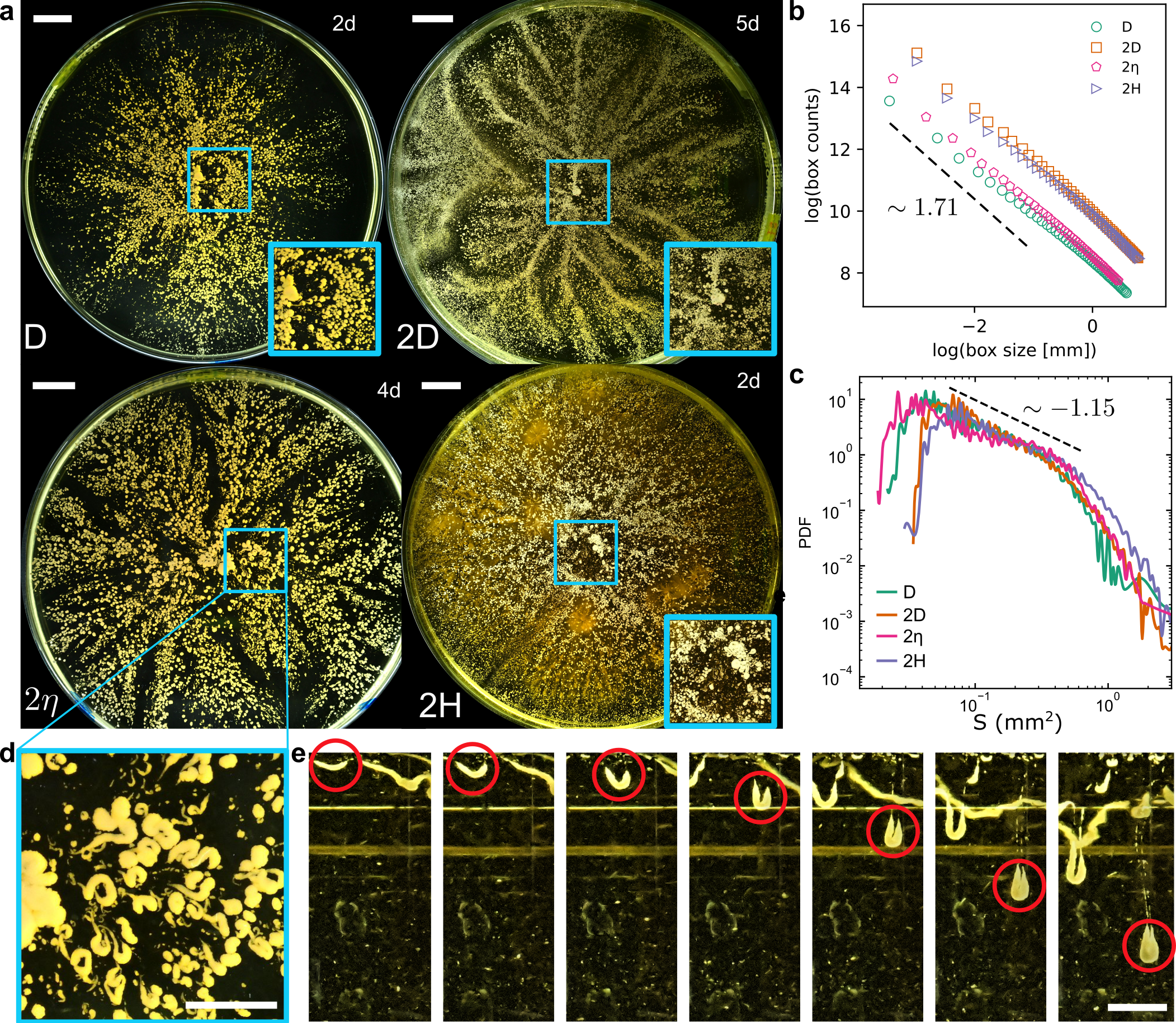}
    \caption{\textbf{Fractal Morphogenesis and the Transport Cycle.} (\textbf{A.}) Diversity of fractal morphologies observed under varying conditions (D: baseline diameter; 2D: double diameter; 2H: double height; 2$\eta$: double viscosity, movies S3-S6). Insets show magnified views. (\textbf{b}) Box-counting analysis reveals a robust fractal dimension $D_{f} \approx 1.71$ across conditions, characteristic of DLA/CDA processes. (\textbf{c}) The probability distribution function (PDF) of satellite colony sizes (S) follows a power law $P(S) \sim S^{-\xi}$, consistent with the hyper-scaling prediction $\xi = 2 - D_f/2$. (\textbf{d}) Close-up image showing characteristic ``horseshoe'' shapes of dispersed seeds. (\textbf{e}) Time sequence (side view) capturing flow-induced stretching, fragmentation, and folding of colony filaments during transport (red circles highlight the formation and detachment process, movie S7, S8). Scale bars, 10~mm (\textbf{a}-left column), 20~mm (\textbf{a}-right column), 5~mm (\textbf{d}), 2.5~mm (\textbf{e}).}
    \label{fig:fig3}
\end{figure}

The same convective flow that drives the explosive kinetics also sculpts the colony morphology, producing intricate fractal patterns across experimental conditions (Fig.~\ref{fig:fig3}\textbf{a}). Despite variations in diameter, height, and viscosity, all patterns exhibited consistent fractal scaling with dimension $D_f \approx 1.71$ (Fig.~\ref{fig:fig3}\textbf{b}), characteristic of Laplacian growth processes such as Diffusion-Limited Aggregation (DLA)~\cite{witten1981diffusion}. This morphogenesis arises from a flow-focusing instability that we term Circulation-Driven Aggregation (CDA), where convergent bottom flow directs settling seeds toward existing biomass while hydrodynamic screening suppresses growth in voids. While DLA is governed by the Laplacian diffusion field, CDA is governed by the incompressible Stokes flow, suggesting a distinct universality class for pattern formation in active, flow-generating systems~\cite{tarboton1988fractal, sreenivasan1991fractals, matsuyama1993fractal, briggs1992fractals}.

The fractal organization produces distinct statistical signatures, with satellite colony sizes following a scale-free distribution $P(S) \sim S^{-\xi}$ where $\xi \approx 1.15$ (Fig.~\ref{fig:fig3}\textbf{c}), consistent with, $\xi = 2 - D_f/2$, our hyperscaling prediction~\cite{mori2020common} (Materials and Methods; Theoretical Framework in Supplementary Material). At the microscopic level, seeds undergo shape transformations, forming characteristic ``horseshoe'' structures through flow-induced stretching and folding (Fig.~\ref{fig:fig3}~\textbf{D, E}; fig.~\ref{figS:Smoke}, movies S7, S8).

\begin{figure}
    \centering
    \includegraphics[width=0.8\textwidth]{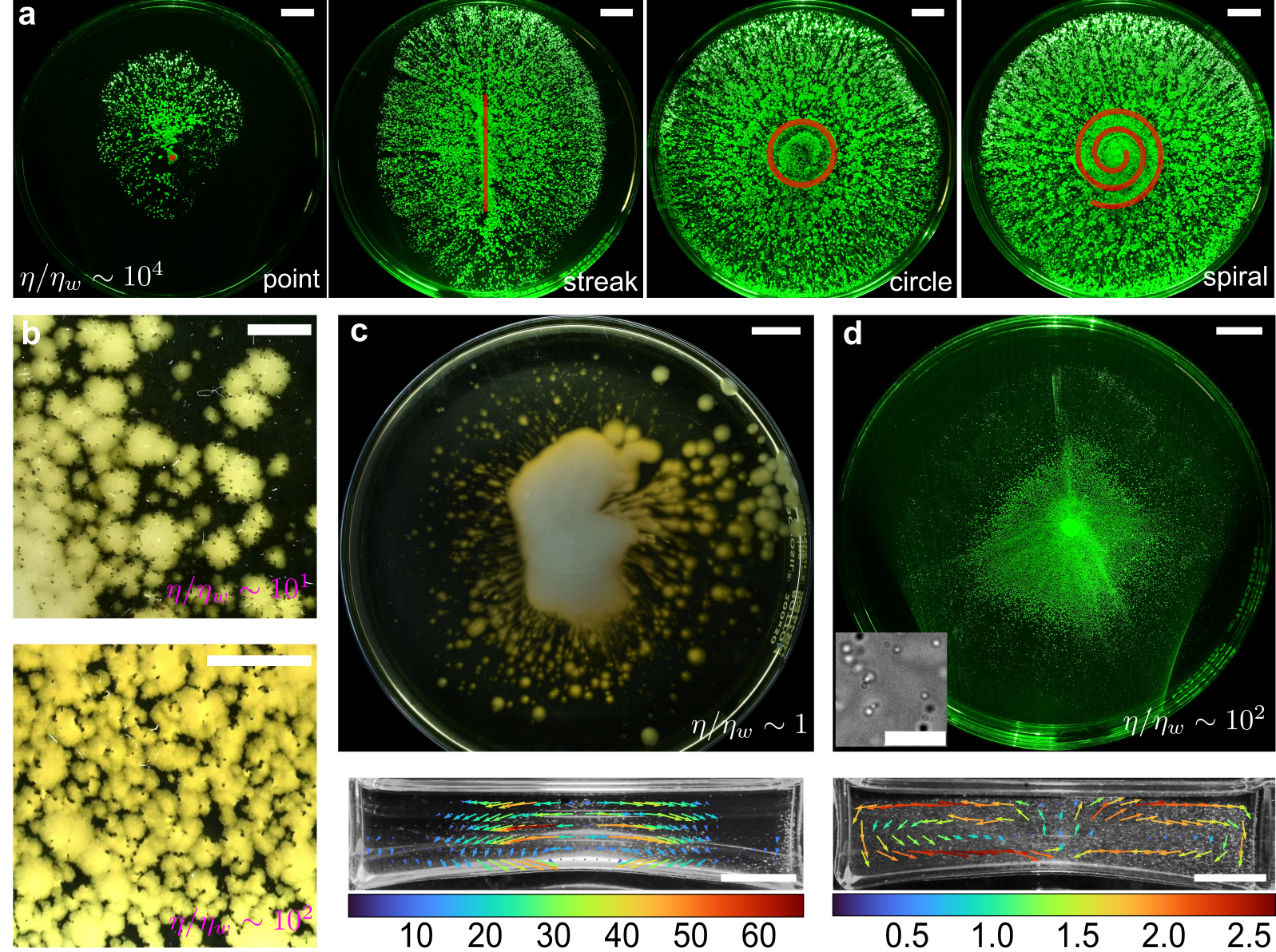}
    \caption{\textbf{Generality of the mechanism.} (\textbf{a}) Robustness to initial conditions. The self-organized dispersal pattern emerges regardless of the initial inoculum geometry (point, streak, circle, spiral) in high viscosity media ($\eta/\eta_{w}\sim10^{4}$). (\textbf{b}) Dispersal persists at medium to low viscosities ($\eta/\eta_{w} \sim 10^2, 10^1$), resulting in ``Swiss-cheese''or ``tafoni''-like morphology due to increased erosion (movie S9).  (\textbf{c}) Dispersal even in nutritious water (YPD, $\eta/\eta_{w} \sim 1$) Bottom: Corresponding PIV flow field (color bar units: mm/h). (\textbf{d}) Fractal colony morphologies observed in a phylogenetically distinct, non-motile coccus microbe at intermediate viscosity ($\eta/\eta_{w} \sim 10^{2}$). Dispersal dynamics (top) and PIV flow field (bottom; color bar units: mm/h) for the coccus microbe, demonstrating the generality of the physics-based strategy.  Scale bars, 10~mm (\textbf{a}), 5~mm (\textbf{b}), 20~mm (\textbf{c}-top), 10~mm (\textbf{c}-bottom), 10~mm (\textbf{d}), 10~$\mu m$ (\textbf{d}-inset), 10~mm (\textbf{d}-bottom).}
    \label{fig:fig4}
\end{figure}

We next tested the generality of the mechanism across different conditions. Strikingly, even when inoculated with only a few yeast cells, the fireworks still took off (fig.~\ref{figS:Intial-Inoc Conditions}\textbf{f}). The fractal fireworks pattern emerged robustly regardless of initial inoculum geometry~\cite{serra2023mechanochemical, everson1992accumulation, meunier2003vortices} (Fig.~\ref{fig:fig4}\textbf{a}; fig.~\ref{figS:Intial-Inoc Conditions})--patterns reminiscent of persistent Lagrangian vortical elements~\cite{haller2015lagrangian, david2018kinematic}--also persisting across viscosity regimes, though with altered ``Swiss-cheese'' or ``tafoni''-like~\cite{groom2015defining} seed morphology at lower viscosities due to increased erosion (Fig.~\ref{fig:fig4}\textbf{b}; movie S9, fig.~\ref{figS:eta-variation}), suggesting breakup instabilities that are different from the Rayleigh-Plateau-like~\cite{shi1994cascade} pinch-off occurring at higher viscosities (Fig.~\ref{fig:fig1}\textbf{e}). The mechanism remained operative even in nutrient-rich water across a wide range of conditions, exhibiting similar circulatory and hopping dynamics. ($\eta/\eta_w \sim 1$; Fig.~\ref{fig:fig4}\textbf{c}, fig.~\ref{figS:YPD-all}). Crucially, the phylogenetically distinct bacterium \textit{Staphylococcus aureus} exhibited identical dispersal dynamics and flow generation (Fig.~\ref{fig:fig4}\textbf{d}; fig.~\ref{figS:Cocci-all}), confirming metabolic fireworks as a general physics-based colonization strategy rather than a species-specific phenomenon.

The `metabolic fireworks' mechanism establishes a new class of non-equilibrium physical systems: a \textit{living} solutal Rayleigh-B\'enard convection~\cite{egolf2000mechanisms} driven by proliferating active matter~\cite{hallatschek2023proliferating}. Its autocatalytic power is a new biophysical finding, coupling the self-generated flow to the \textit{frangible} mechanics of the colony. This use of fragmentation to create new, \textit{proliferating} sources is a biological analogue to autocatalytic chemical plumes~\cite{rogers2005buoyant}, but replaces a fluid-dynamical `pinch-off' with a biomechanical fracture, driving the observed explosive kinetics. The simplicity of the underlying physics—buoyancy-driven flow coupled with fragmentation—suggests broad generality. The `metabolic engine' itself is not restrictive and could be driven by thermal or effervescent convection. This mechanism provides a fundamental physical strategy for niche construction~\cite{thompson1992growth}, with broad implications for biofilm sloughing, microbial ecology in sediments~\cite{bar2018biomass, pan2013structure}.

\bibliography{04_references}

\clearpage
\newpage
\section*{Methods}

\subsection{Experimental Methods}

\subsubsection{Strains and Culture Conditions}
\label{methods:strains-culture}
Experiments were primarily conducted using the yeast \textit{Saccharomyces cerevisiae} (Strain: CEN.PK). A phylogenetically distinct, non-motile coccus bacterium, \textit{Staphylococcus aureus} was also used for generality experiments (Main Text Fig.~\ref{fig:fig4}\textbf{d}). Yeast/Cocci cultures were grown overnight in 10 mL of standard YPD/LB (Yeast Extract-Peptone-Dextrose)/(Lysogeny Broth) medium in 50 mL falcon tubes. All experiments were incubated at 30$^\circ$C.

\subsubsection{Preparation of Viscous Media}
\label{methods:HEC-prep}
To create the quiescent, viscous environment, the YPD medium (containing 1\% yeast extract, 2\% peptone, 2\% D-(+)-glucose; G8270, Sigma-Aldrich) was supplemented with 2-Hydroxyethyl cellulose (HEC, 434981-Sigma-Aldrich) in concentrations ranging from 0.2\% to 1.6\% (w/v). The HEC powder was dissolved by stirring the solutions for 3 hours at 200 rpm using an IKA-RCT Digital magnetic stirrer. The solution was then sterilized by microwaving for 2 minutes at a medium-high setting (protocol similar to~\cite{atis2019microbial}). The sterilized media was left overnight to cool to room temperature and allow bubbles to dissipate before use the next day (for standard experiments and rheology measurements). The resulting viscosities ranged from approximately $10^1$ to $2\times10^4$ times the viscosity of water (Table~\ref{tab:EC-eta-details}). The density of the viscous fluid substrate is close to that of water in all scenarios \citep{atis2019microbial, weinstein2018microbial}. 

\subsubsection{Rheology measurements and viscosity values}
\label{methods:rheology-viscosity}

Rheological measurements were performed using an Anton-Paar rheometer (MCR 302e) equipped with a cone–plate geometry (CP50-4/S, diameter = 50~mm, angle = 4$^o$), over a shear rate range of $0.0001$–$500~s^{-1}$~(Supplementary Fig.~\ref{figS:Rheometry}). 

The shear rate was estimated as $\dot{\gamma} \sim U_{p}/H$ (Fig.~\ref{fig:fig1}\textbf{b}, \ref{figS:eta-variation}\textbf{l}). For $\eta/\eta_w \sim 10^4$, this gives $\dot{\gamma} \sim 7\times10^{-5}$~s$^{-1}$, and for $\eta/\eta_w \sim 10^2$, about $\dot{\gamma} \sim 5\times10^{-4}$~s$^{-1}$ for H=8~mm layers. Both values lie below the measurable range of the rheometer as seen in Fig.~\ref{figS:Rheometry}, making direct viscosity estimation difficult. Therefore, representative viscosities values were assigned from the plateau regions of the flow curves corresponding to each polymer concentration (tabulated in \ref{tab:EC-eta-details}).

\begin{table}[htbp]
\centering

\begin{tabular}{@{}lll@{}} 
\toprule
\% HEC (w/v) & Ratio ($\eta/\eta_{\text{water}}$) & $\eta$ (Pa$\cdot$s) \\
\midrule
0.2\%          & $\sim 10^1$             & $\approx 0.01$ \\
0.4\%          & $\sim 10^2$             & $\approx 0.1$ \\
0.8\%          & $\sim 2\times 10^3$     & $\approx 2$ \\
1.2\%          & $\sim 10^4$             & $\approx 10$ \\
1.6\%          & $\sim 2\times 10^4$     & $\approx 20$ \\
\bottomrule
\end{tabular}
\caption{Viscosity of HEC-supplemented media. Absolute viscosity $\eta$ can be estimated assuming $\eta_{\text{water}} \approx 0.001$ Pa$\cdot$s, or measured via rheometry (Fig.~\ref{figS:Rheometry}).}
\label{tab:EC-eta-details}
\end{table}

\subsubsection{Smoke visualization}
\label{methods:smokeviz}
To ascertain the nature of the flow, we performed a simple analog experiment using a rising smoke plume~(Supplementary Fig.~\ref{figS:Smoke}). Incense sticks were placed at the bottom center of a square container ($L = 200$~mm, $H = 30$~mm) with the top and sides enclosed (Fig.~\ref{figS:Smoke}). The smoke rose upward, struck the ceiling, spread radially outward, and then turned downward as it approached the side edges. Distinct horseshoe-shaped structure became visible during this descent—remarkably similar to those observed in viscous yeast experiments—indicating Rayleigh–Bénard-type convective circulations.

\subsubsection{Experimental Setup and Inoculation}
\label{methods:setup-inoculation}
Experiments were performed in sterile glass containers (e.g., petri dishes of varying diameters and heights). The containers were housed in a custom-built incubator maintained at 30$^\circ$C under clean conditions. 

To prepare the inoculum, the overnight culture was centrifuged for 3 minutes at a speed of 3 rcf. 9.9 mL of the supernatant was discarded, and the remaining pellet was resuspended in the residual 0.1 mL to create a dense yeast slurry.

For standard experiments (point inoculation), 2.5 $\mu$L of this slurry was inoculated at the bottom center of the container. The pipette tip was inserted through the viscous medium to the bottom glass surface, the inoculum was dispensed, and the pipette was vertically retracted. Retraction was performed rapidly to minimize the formation of a vertical trail, which typically settled over the inoculum or was occasionally entrained by the initial circulatory loops.

For experiments testing the effect of initial geometry (Main Text Fig.~\ref{fig:fig4}\textbf{a}), a modified inoculum was prepared: 10 mL of overnight culture was centrifuged, the supernatant was discarded, and the pellet was resuspended in 1 mL of fresh YPD (a 10x dilution compared to the standard slurry). 10 $\mu$L of this diluted solution was used to manually draw the desired shapes (line-streak, circle, spiral) at the bottom of the container.

After pouring the liquid substrates, the Petri dishes were sealed with lids and wrapped with Parafilm to minimize evaporation. Evaporation was monitored by weighing the containers before and after the experiment; typical losses were 0.5–1 g per day for standard 90 mm Petri dishes (approximately 1–3 \% of the substrate weight). Typical layer heights were maintained at $H = 8$~mm, using dishes of diameter $D = 90$~mm.

\subsubsection{Imaging and Analysis}
\label{methods:imaging-analysis}

The colonization dynamics were typically imaged from directly underneath the container using a DSLR camera (Nikon Z-series). Time-lapse images were acquired at 10-20 minute intervals for 2-4 days. 

Image analysis was performed using custom Python scripts. The colony area, $A(t)$, was estimated using a uniform intensity threshold across all images and experiments. Power-law fitting was applied to the kinetic data during the explosive growth phase, within selected time windows, to obtain the area growth exponent following $A(t) \sim t^{\alpha}$. The fitting typically yielded a mean absolute percentage error $MAPE \sim <10\%$ and an $R^2 > 0.95$, ensuring reliable exponents. Larger dishes (diameter $D = 180,\mathrm{mm}$) were used specifically to enable longer-term growth and to allow the colonies to expand without geometric confinement.

The crown expansion diameter, $2R(t)$, was determined by manually marking a multipoint polygon along the faint outer boundary of the expanding crown (Fig.~\ref{figS:Controls-extras}\textbf{c}). From these contours, a mean diameter was computed to extract the scaling exponent $2R(t) \sim t^{\beta}$.

Growth of individual seed clusters, $S(t) \sim t^{\gamma}$, was tracked manually for selected isolated clusters positioned at distinct angular locations—chosen specifically to avoid overlap or merging with neighboring clusters (Fig.~\ref{figS:Controls-extras}\textbf{a,b}). Each seed was followed from its initial faint loop-out phase through to the landing time ($\tau_{\text{loop}}$; Fig.~\ref{figS:4cases-prt1}\textbf{c}) and subsequent saturation. This progression corresponds to the local area growth saturating at various angular positions as the radially propagating “fireworks” fronts pass through them (Fig.~\ref{figS:Controls-extras}\textbf{d}).

Particle Image Velocimetry (PIV) was used to quantify the flow fields. A 5W-532$nm$ continuous laser in sheet-mode was used to illuminate the flow-field. Nearly neutrally buoyant 20$\mu m $ polyamide particles was used to seed the flow (LaVision) and the PIVLAB software was used to estimate the flow-fields. The resulting time-series flow fields were checked for outliers and averaged over suitable time windows (depending on flow speeds and viscosities) to obtain smooth vector-field representations.

\subsubsection{Fractal Dimension}
\label{methods:frac-dim-calc}
The ImageJ plugin FracLac was used to obtain a robust estimate of the fractal dimension, $D_f$. Unlike standard box-counting algorithms, FracLac employs shifting grids to provide a more reliable estimate of $D_f$. Box sizes ranging from 1 to 50 pixels were employed, with the maximum size set to approximately twice the diameter of the largest clusters identified in the cluster size analysis. FracLac outputs counts versus box-size/image-width, and when plotted on a log–log scale, the slope of the resulting line corresponds to the fractal dimension.

\subsubsection{Cluster size distribution}
\label{methods:cluster-psd}
Images were converted to binary, and the ImageJ plugin MorphoLibJ was used to perform distance-transform watershed segmentation, enabling the identification and segmentation of individual clusters. The plugin was then used to calculate cluster centroids and areas, which were subsequently used to determine the nearest-neighbour distances and cluster size distributions (Supplementary Fig.~\ref{figS:4-cases-extras2}). 

\subsubsection{Non-dimensional numbers}
\label{methods:non-dim-numbers}

The Reynolds and Rayleigh numbers can be expressed in terms of the plume velocity, $1 < U_p < 100$~mm~h$^{-1}$, roughly estimated from the PIV velocity measurements (Fig.~\ref{fig:fig1}\textbf{b}, \ref{figS:eta-variation}\textbf{l}, \ref{figS:YPD-all}\textbf{b}). The Reynolds number is $Re = \rho U_p H / \eta$, where $\rho$, $\eta$ is the density and dynamic viscosity of the fluid, and the Rayleigh number is $Ra = U_p H / \alpha_D$, where $\alpha_D$ is the solute diffusivity ($\alpha_D \sim 10^{-10} m^2/s$ in YPD, \cite{atis2019microbial}). The corresponding ranges are $Re \sim 10^{-1}$–$10^{-7}$ for velocities of $100$–$1$~mm~h$^{-1}$ (from YPD to highly viscous cases, $\eta/\eta_w \sim 1$–$10^4$) and $Ra \sim 10^{3}$–$10^{5}$. The classical critical $Ra_c \simeq 1700$ applies to convection driven by uniform heating across a thin fluid layer, whereas here the flow originates from a localized point inoculum—effectively a point heat source—that gradually expands as an area-source. Thus, a direct one-to-one comparison of critical thresholds may be only roughly appropriate. Table~\ref{tab:dimless} summarizes the relevant dimensionless numbers characterizing the physical regime.

\begin{table}[h!]
\centering
\small
\begin{ruledtabular}
\begin{tabular}{lcc}
Quantity & Definition & Estimate \\
\hline
Reynolds number
& $\mathrm{Re} = \rho U_p H / \eta$
& $10^{-7}$--$ 10^{-1}$ \\
P\'eclet number
& $\mathrm{Pe} = U_p R / \alpha_D$
& $10^{6}$--$10^{4}$ \\
Damk\"ohler-like ratio
& $\mathrm{Da} = (R/\dot{R})/(R/U_p) = U_p/\dot{R}$
& $1$--$10$ \\
Effective Rayleigh number
& $\mathrm{Ra}_{\text{eff}} = U_p H / \alpha_D$
& $10^{5}$--$ 10^{3}$ \\
\end{tabular}
\end{ruledtabular}
\caption{Key dimensionless numbers characterizing the viscous / convective regime. Estimates use representative experimental values: $U_p \simeq 1$--$100$ mm h$^{-1}$, $R \sim 45$~mm, $H = 8$~mm, $\alpha_D \simeq 10^{-10}$ m$^{2}$ s$^{-1}$ (in YPD, $\alpha_D \sim 1/\eta$), $\rho_0 \simeq 10^{3}$ kg m$^{-3}$, $\eta \sim 10^{-3}$--$20$ Pa s.}
\label{tab:dimless}
\end{table}

In the most viscous / shallow configurations (large $\mu$, small $H$) the classical form of $\mathrm{Ra}_{\text{eff}}$ can drop to $\mathcal{O}(10^{2})$, but in the parameter range used to extract the scaling exponents (Fig.~\ref{fig:fig2}\textbf{h}), it lies in the range $10^{4} < Ra_{eff}<10^{5}$, i.e. safely within the convective regime. All reported $\mathrm{Pe}$ are computed from the measured $U_p$ and colony or 'firework' scale $R$ in the same experiments that yield $\alpha$ and $\beta$. In the main text we primarily report $\mathrm{Re}$, $\mathrm{Pe}$ and $\mathrm{Da}$ computed from estimated $U_p$ and the set values of $R$ and $H$ for each run. The $Ra_{eff}$, an experimentally convenient form estimated using the measured plume or recirculation speed $U_p$, is used only to indicate that the system operates safely within the convective, solutal Rayleigh B\'enard regime.

\subsection{Theoretical Framework}
\label{methods-si:theory}

This theoretical framework assumes, in the regime of the present experiments: (i) a thin-gap Stokes/Hele–Shaw flow with aspect ratio $\varepsilon = H/R \ll 1$ and low Reynolds number $\mathrm{Re} = U_p H/\nu \ll 1$; (ii) high P\'eclet number $\mathrm{Pe} = U_p R / \alpha_D \gg 1$ so that advection dominates lateral solute transport over the scale of the colony; (iii) the Boussinesq approximation with $|\Delta\rho|/\rho_0 \ll 1$ so that density variations only enter as a buoyancy term; and (iv) quasi-steady metabolism, i.e. the metabolic production rate per unit biomass $s_0$ varies on timescales longer than the lateral advective time $R/U_p$. Under these assumptions, the same mechanism (metabolically generated buoyancy driving toroidal circulation in a confined layer) links the flow amplitude to the areal biomass and thereby to the observed explosive growth laws. Below we first derive the 2D, depth-averaged scaling that gives the parameter-free relation $\beta = (\alpha+2)/4$, and then generalize it to an effective (fractally thickened) biomass distribution.

\subsubsection{Growth Scaling: 2D Homogeneous Model}
\label{methods-si:growthScaling}

We consider a liquid layer of height $H$ above a horizontal substrate. Microbial metabolism locally produces solute concentration anomalies $c(\mathbf{x},t)$, which in turn induce density anomalies
\[
\rho' = -\rho_0 \beta_s c,
\]
where $\beta_s$ is the solutal expansion coefficient appropriate to the metabolically produced solute(s). Using a standard thin-gap depth-average, the incompressible flow $\mathbf{u}(x,y,t)$ obeys the Hele–Shaw / Darcy form
\begin{equation}
\frac{12\mu}{H^{2}} \,\mathbf{u} = -\nabla p + \rho' g \,\hat{\mathbf{z}}, 
\qquad 
\nabla \cdot \mathbf{u} = 0,
\end{equation}
with negligible inertia ($\mathrm{Re} \ll 1$). Localized buoyancy over a vertical extent $H$ produces a lateral pressure variation of order
\[
\nabla p \sim \frac{g H\, \Delta\rho}{R}
\]
over the colony scale $R$. Balancing this with the Darcy-like viscous resistance gives
\begin{equation}
U_p \sim \frac{H^{2}}{12\mu} \,\frac{gH}{R} \Delta\rho 
= C_{\mu} \frac{g H^{3}}{\mu} \frac{\Delta\rho}{R},
\label{eq:U-darcy}
\end{equation}
where $C_{\mu}$ is a geometric factor.

The solute field obeys a depth-averaged advection–diffusion–source equation:
\begin{equation}
\partial_{t} c + \mathbf{u} \cdot \nabla c 
= \alpha_D \nabla^{2} c + s_0 \,\phi(t)\, \chi_A,
\label{eq:adv-diff}
\end{equation}
where $\alpha_D$ is the solute diffusivity, $s_0$ is the solute production rate per unit projected biomass, $\phi(t)$ is the dimensionless biomass area density, and $\chi_A$ localizes production to the colony footprint.

\textit{Non-dimensionalization and ordering.} Let $x = R \tilde{x}$, $t = (R/U_p) \tilde{t}$, $\mathbf{u} = U_p \tilde{\mathbf{u}}$, $c = c_{\star} \tilde{c}$, where $c_{\star}$ is a characteristic concentration anomaly. Equation~\eqref{eq:adv-diff} becomes
\begin{equation}
\mathrm{Pe} \,\tilde{\mathbf{u}} \cdot \tilde{\nabla} \tilde{c} 
= \tilde{\nabla}^{2} \tilde{c} 
+ \Lambda \,\phi \,\tilde{\chi}_{A} 
+ O\!\left(\frac{1}{\mathrm{Da}} \partial_{\tilde{t}} \tilde{c}\right),
\quad
\mathrm{Pe} = \frac{U_p R}{\alpha_D}, 
\quad
\Lambda = \frac{s_0 R^{2}}{\alpha_D c_{\star}},
\label{eq:nondim}
\end{equation}
where we have grouped the unsteady term with a Damk\"ohler-like factor
\[
\mathrm{Da} 
= \frac{\text{front-evolution time}}{\text{advective time}}
= \frac{R / \dot R}{R / U_p}
= \frac{U_p}{\dot R}
= \frac{U_p}{\beta R / t}.
\]
In the experiments, measured velocities and front speeds give $t_a = R/U_p \ll t_f = R/\dot R$ over the explosive window, so $\mathrm{Da} \geq 1$, and from measured $U_p$, $R$ and an estimated $\alpha_D$ for the viscous media, $\mathrm{Pe} \gg 1$ as well. Thus, in the source region the leading balance is advective transport of solute against its metabolic supply:
\begin{equation}
\tilde{\mathbf{u}} \cdot \tilde{\nabla} \tilde{c} 
\sim \frac{\Lambda}{\mathrm{Pe}} \phi
\quad \Rightarrow \quad
\frac{U_p \,\Delta c}{R} \sim s_0 \phi
\quad \Rightarrow \quad
\Delta c \sim \frac{s_0 \phi R}{U_p}.
\label{eq:delta-c}
\end{equation}

With $\Delta\rho = \rho_0 \beta_s \Delta c$ and inserting \eqref{eq:delta-c} into \eqref{eq:U-darcy}, we obtain the closure
\begin{equation}
U_p \sim C_{\mu} \frac{g H^{3}}{\mu} \frac{1}{R} \,\rho_0 \beta_s \frac{s_0 \phi R}{U_p}
\quad \Rightarrow \quad
U_p^{2} \sim \left(C_{\mu} \rho_0 \beta_s s_0 \frac{g H^{3}}{\mu}\right) \,\phi.
\label{eq:U2-phi}
\end{equation}
All dimensional prefactors collect into the bracketed constant; the \emph{only} dependence on the evolving biomass enters through $\phi$. Thus
\begin{equation}
U_p \propto \phi^{1/2}.
\label{eq:U-phiHalf}
\end{equation}

Since $\phi = A/R^{2}$ for a homogeneous layer,
\begin{equation}
\dot{R} \sim U_p \propto \left(\frac{A}{R^{2}}\right)^{1/2}
\quad \Rightarrow \quad
\dot{R} \propto A^{1/2} R^{-1}.
\label{eq:Rdot}
\end{equation}
Assuming power laws $A(t) \sim t^{\alpha}$ and $R(t) \sim t^{\beta}$, we have
\[
t^{\beta-1} \propto t^{\alpha/2} t^{-\beta}
\quad \Rightarrow \quad
\beta - 1 = \frac{\alpha}{2} - \beta
\quad \Rightarrow \quad
\boxed{\beta = \frac{\alpha + 2}{4}}.
\]
Remarkably, this relationship derived from first principles accurately captures the experimental data across all conditions tested (Fig.~\ref{fig:fig2}\textbf{h}), despite making the simplifying assumption of homogeneous biomass distribution.

\subsubsection{Incorporating Fractal Morphology and 3D Distribution}
\label{methods-si-theory:fractal-3d}

The success of the homogeneous model is initially surprising given the clearly fractal nature of colony patterns observed experimentally ($D_f \approx 1.71$). To resolve this apparent paradox, we now incorporate the effects of fractal geometry and three-dimensional biomass distribution.\\\par

The key insight is that metabolically active biomass is distributed throughout the fluid column, not merely confined to the bottom surface (Fig.~\ref{fig:fig1}\textbf{f}). While the colony exhibits fractal morphology in 2D projection, the effective biomass relevant for flow generation scales with a different dimension due to this 3D distribution.\\\par

We generalize the biomass density to account for fractal scaling:

\begin{equation}
\phi_{\text{eff}} \propto \frac{A_{\text{eff}}}{R^{2}} \propto R^{D_{\text{eff}} - 2},
\end{equation}
where $A_{\text{eff}} \propto R^{D_{\text{eff}}}$ is the metabolically active content, and $D_{\text{eff}}$ accounts for both the planar fractal geometry and the vertical extent of active material. Repeating the scaling from \eqref{eq:U-phiHalf} with $\phi$ replaced by $\phi_{\text{eff}}$ gives
\begin{equation}
U_p \propto \phi_{\text{eff}}^{1/2} \propto R^{\frac{D_{\text{eff}} - 2}{2}}.
\end{equation}
Since $U_p = dR/dt \propto t^{\beta - 1}$ and $R \propto t^{\beta}$, we obtain
\begin{equation}
t^{\beta - 1} \propto t^{\beta \frac{D_{\text{eff}} - 2}{2}}
\quad \Rightarrow \quad
\beta - 1 = \beta \frac{D_{\text{eff}} - 2}{2}.
\end{equation}
Solving for the exponents gives
\begin{equation}
\beta = \frac{2}{4 - D_{\text{eff}}}, 
\qquad
\alpha = \beta D_{\text{eff}} = \frac{2 D_{\text{eff}}}{4 - D_{\text{eff}}}.
\label{eq:Deff-rel}
\end{equation}

Equation~\eqref{eq:Deff-rel} has two useful interpretations. First, the homogeneous, depth-averaged case corresponds to $D_{\text{eff}} = 2$ and immediately recovers $\beta = (\alpha+2)/4$. Second, allowing $D_{\text{eff}} > 2$ naturally generates exponent pairs in the \emph{observed} range. For example, taking the measured front exponent $\beta \simeq 1.30$ gives
\[
D_{\text{eff}} = 4 - \frac{2}{\beta} \simeq 4 - \frac{2}{1.30} \simeq 2.46,
\]
and then $\alpha = \beta D_{\text{eff}} \simeq 1.30 \times 2.46 \simeq 3.2$, i.e. within the experimental scatter of the area exponents ($\alpha \approx 3.1$–$3.5$ across depths and viscosities). We therefore regard the $D_{\text{eff}}$-extension as a descriptive refinement that explains the \emph{spread} of exponents around the 2D law, not as a second, independent constraint. The collapse onto $\beta = (\alpha+2)/4$ is the primary result; the effective-dimension view shows that modest thickening of the active set ($D_{\text{eff}} \approx 2.3$–$2.5$) is sufficient to account for the deviations.

\subsubsection{Morphogenesis: Particle Size Distribution}
\label{methods-si:linking-xi-to-Df}

We derive the link between the fractal geometry ($D_f$) and the size distribution exponent ($\xi$) using a hyperscaling argument. The total colonized area $A_{\text{total}}$ within radius $R$ scales as $A_{\text{total}} \sim R^{D_f}$. The same area can be written as an integral over cluster sizes:
\begin{equation}
A_{\text{total}} = \int^{S_{\text{max}}} S' \, P(S') \, dS'.
\end{equation}
We assume that individual yeast clusters are dense ($d=2$) and that the largest cluster size is set by the system size, $S_{\text{max}} \sim R^{2}$. Taking $P(S) \sim S^{-\xi}$ and using that $\xi<2$ in the experiments, the integral is dominated by the upper cutoff:
\begin{equation}
A_{\text{total}} \sim S_{\text{max}}^{2-\xi} \sim R^{2(2-\xi)} = R^{4 - 2\xi}.
\end{equation}
Equating the two expressions for $A_{\text{total}}$ gives
\begin{equation}
\xi = 2 - \frac{D_f}{2}.
\end{equation}
For $D_f \approx 1.71$, this predicts $\xi \approx 1.145$, in good agreement with the measured size distributions.

\subsubsection{Kinetic Constraint: Local Growth and Seeding Convolution}
\label{methods-si:seeding-convolution-model}

Explosive, system-level growth arises from the interplay between (i) local growth of each deposited fragment and (ii) the rate at which flow-driven fragmentation generates new seeds. Let $S(t) \sim t^{\gamma}$ be the local growth law of an individual colony and let the seeding rate scale as $dN/dt \sim t^{\gamma_s}$. The total colonized area is then the convolution
\begin{equation}
A(t) = \int_{0}^{t} \frac{dN}{dt'}(t') \, S(t - t') \, dt' 
\;\propto\; \int_{0}^{t} (t')^{\gamma_s} (t - t')^{\gamma} \, dt'.
\end{equation}
With the change of variables $u = t'/t$,
\begin{equation}
A(t) \propto t^{\gamma_s + \gamma + 1} \int_{0}^{1} u^{\gamma_s} (1-u)^{\gamma} \, du
\;\propto\; t^{\gamma_s + \gamma + 1},
\end{equation}
since the Beta-function integral is a constant prefactor. Identifying $A(t) \sim t^{\alpha}$ gives the kinetic constraint
\begin{equation}
\boxed{\alpha = \gamma + \gamma_s + 1.}
\end{equation}
Thus, for the experimentally measured global exponent $\alpha \gtrsim 3$ and the local growth exponent ($\gamma \approx 1.0$–$1.9$) we obtain a superlinear seeding exponent $\gamma_s \gtrsim 1$ (Supplementary Fig.~\ref{figS:4cases-prt1}\textbf{c}, showing correlation between $\alpha$, $\gamma_s$ and $N_{p}$-Number of clusters at 2 days).



\subsection*{Acknowledgements} 

We acknowledge the Mechanical Workshop Facility at the NCBS, Bangalore. We acknowledge support from the Department of Atomic Energy (India), under project no.\,RTI4006, and the Simons Foundation (Grant No.\,287975). 

\subsection*{Author contributions} 

J.D. and S.T. conceived the project. J.D. discovered the phenomenon, designed and performed the experiments, and analysed the data. J.D. and S.T. developed the theoretical framework. J.D. and S.T. wrote the paper.

\subsection*{Competing interests} 
The authors declare no competing interests.
\setcounter{figure}{0}
\setcounter{section}{0}
\renewcommand{\figurename}{Fig.}
\renewcommand{\thefigure}{S\arabic{figure}}
\renewcommand{\thetable}{S\arabic{table}}
\renewcommand{\thesection}{S\arabic{section}}
\renewcommand{\thesubsection}{\thesection.\arabic{subsection}}

\clearpage

{\centering \huge \textbf{Supplementary Information}}

\section{Supplementary Multimedia}
\label{si:media}

\textbf{Movie 1:} Development of the metabolic fireworks in side view (Main Text Fig.~\ref{fig:fig1}\textbf{a}). \\\par
\textbf{Movie 2:} Fragmentation of clusters within the circulatory loop (Main Text Fig.~\ref{fig:fig1}\textbf{e}). \\\par
\textbf{Movie 3:} Time-lapse of the metabolic fireworks phenomenon (bottom view), showing explosive kinetics and anisotropic, channeled expansion (Main Text Fig.~\ref{fig:fig1}\textbf{f}, Fig.~\ref{fig:fig2}\textbf{b}, Fig.~\ref{fig:fig3}\textbf{a}, case–2D; $H = 8$~mm, $D = 180$~mm, $\eta/\eta_w \sim 10^4$). \\\par
\textbf{Movie 4:} Development of metabolic fireworks for the viscous case ($H = 8$~mm, $D = 90$~mm, $\eta/\eta_w \sim 10^4$; Main Text Fig.~\ref{fig:fig3}\textbf{a}, case–D). \\\par
\textbf{Movie 5:} Development of metabolic fireworks for a higher viscosity case ($H = 8$~mm, $D = 90$~mm, $\eta/\eta_w \sim 2\times10^4$; Main Text Fig.~\ref{fig:fig3}\textbf{a}, case–2$\eta$). \\\par
\textbf{Movie 6:} Development of metabolic fireworks at increased layer height ($H = 16$~mm, $D = 180$~mm, $\eta/\eta_w \sim 10^4$; Main Text Fig.~\ref{fig:fig3}\textbf{a}, case–2H). \\\par
\textbf{Movie 7:} Formation of horse-shoe structures (Main Text Fig.~\ref{fig:fig3}\textbf{e}). \\\par
\textbf{Movie 8:} Smoke visualization experiment (Supplementary Fig.~\ref{figS:Smoke}). \\\par
\textbf{Movie 9:} Fireworks in lower-viscosity substrates (Main Text Fig.~\ref{fig:fig4}\textbf{b}, bottom; $H = 8$~mm, $D = 90$~mm, $\eta/\eta_w \sim 10^2$). \\\par

\section{Supplementary Figures List}
\label{si:figures}
\begin{itemize}
   \item \textbf{Fig.~S1:} Controls with top inoculation in air–viscous substrate interface. Growth in agar plates, stunted growth in low sugar concentrations, and corresponding flow fields at later times and PIV controls .

\item \textbf{Fig.~S2:} Tracking growth within circulatory loops in side and bottom views. Local growth saturation as the firework front passes. Detailed kinetics for the four cases: front speed ($R \sim t^{\beta}$) and local growth ($S \sim t^{\gamma}$).

\item \textbf{Fig.~S3:} Effect of layer height variations ($H = 4$–16~mm) and tilt experiments ($h_{tilt}\approx $2.5~mm). Corresponding views and growth curves.

\item \textbf{Fig.~S4:} The four cases analyzed in detail (D, 2D, 2$\eta$, 2H)—morphology overview at 2~days and corresponding area growth curves ($A \sim t^{\alpha}$).

\item \textbf{Fig.~S5:} Cluster statistics for the four cases: nearest-neighbor spacing ($d_{NN}$) and size distributions ($\tau$). 

\item \textbf{Fig.~S6:} Flow visualizations showing horseshoe formation, smoke plume analogy, and seedling formation in tall–narrow bottle experiments.

\item \textbf{Fig.~S7:} Robustness to initial and inoculation conditions (serial dilutions and lag-phase effects). 

\item \textbf{Fig.~S8:} Views and area growth curves for varying viscosities ($\eta/\eta_w \sim 10^1$–$10^3$) in small and large dishes.

\item \textbf{Fig.~S9:} Experiments with yeast in YPD ($\eta/\eta_w \sim 1$), for a wide-range of conditions.

\item \textbf{Fig.~S10:} Experiments with non-motile cocci bacteria (Staphylococcus aureus) across viscosities ($\eta/\eta_w \sim 1$–$10^4$), in both bottom and top inoculations.

\item \textbf{Fig.~S11:} Rheometry data.
\end{itemize}


\begin{figure*}[!]
    \centering
    \includegraphics[width=\textwidth]{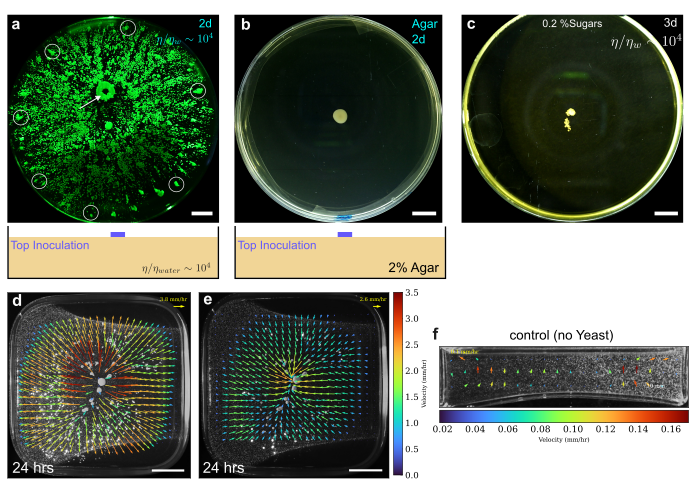}
    \caption{\textbf{Controls.} \textbf{a.} Simultaneous bulk \textit{metabolic firework} and \textit{range expansion} at the air-liquid substrate interface~\cite{atis2019microbial} (white circles) observed for $\eta/\eta_w \sim 10^{4}$, $H = 8$ mm. When yeast is inoculated on the top surface of the liquid substrate (white arrow), part of the inoculum detaches and sinks to the bottom of the dish, initiating the firework, while the remaining portion forms a donut-like top colony (white arrow), leading to a parallel range expansion at the interface. \textbf{b.} Slow radial growth in 2\% agar. \textbf{c.} Very slow dynamics under low-sugar conditions (0.2\%), affirming the role of metabolism driving the flows. \textbf{d,e.} Top and bottom plane PIV flow fields at 24 h. \textbf{f.} Flow-field control in side view without yeast. Scale bars, 10~mm (\textbf{a-f}).}
    \label{figS:Controls}
\end{figure*}


\begin{figure*}[!]
    \centering
    \includegraphics[width=.9\textwidth]{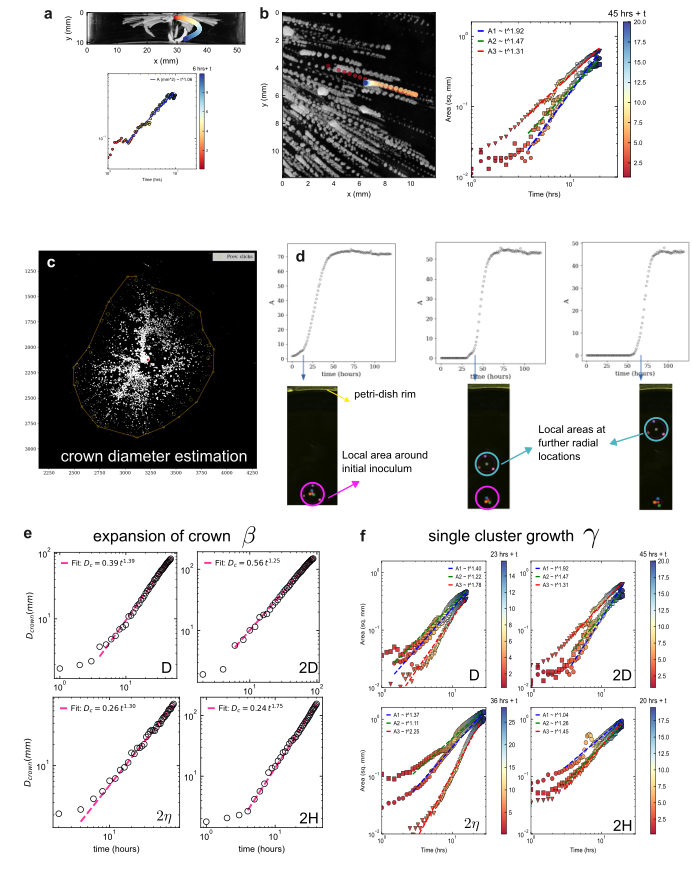}
    \caption{\textbf{Growth kinetics in viscous regime.} \textbf{a.} Side-view area growth of the cluster from the time of initial inoculation (Supplementary Movie 1). \textbf{b.} Bottom-view growth of individual seeds as they move radially outward and return inward for three clusters at different angular positions (Supplementary Movie 3), starting at 45~h (in the power-law phase), with higher cluster growth exponents compared to the initial-time seeds . The background images in \textbf{a} and \textbf{b} are composites generated by adding images over suitable time intervals using FlowTrace~\cite{gilpin2017flowtrace}. \textbf{c.} Manual estimation of the crown diameter by tracing a polygon around the periphery of the firework pattern. \textbf{d.} Growth within a small, selected circular region showing that local expansion saturates over time as the firework crosses these regions. \textbf{e,f.} Radial crown expansion ($2R \sim t^{\beta}$) and single-cluster growth ($S \sim t^{\gamma}$) for variations in height $H$, diameter $D$, and viscosity $\eta$, corresponding to fig.~\ref{fig:fig2}-\textbf{h} and \ref{fig:fig3}-\textbf{a} (Supplementary Movie 3-6).}
    \label{figS:Controls-extras}
\end{figure*}

\begin{figure*}[!]
    \centering
    \includegraphics[width=\textwidth]{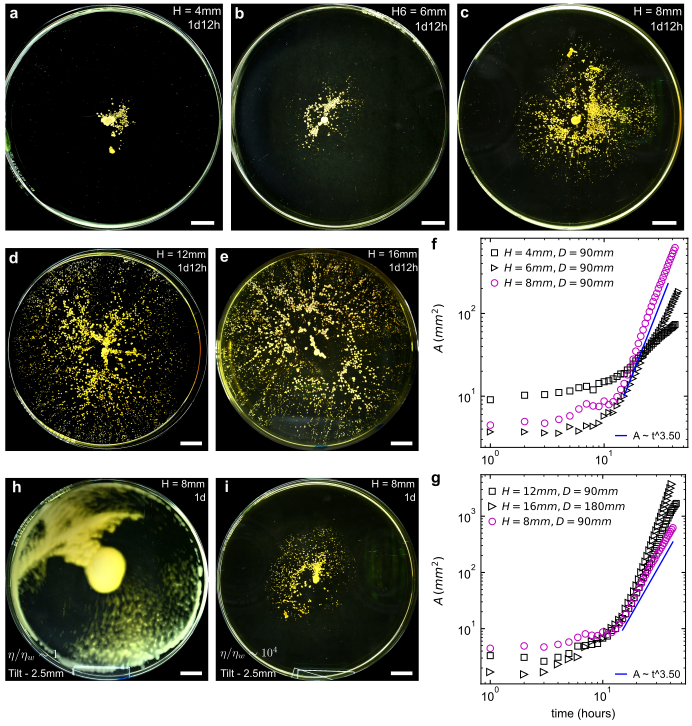}
    \caption{\textbf{Variation in fluid layer height, $H$.} \textbf{a–e.} Progressively faster firework expansions with increasing layer heights ($H = 4$–16~mm) at 1~day~12~h. \textbf{f,g.} Corresponding area growth curves, with reference line $A \sim t^{3.5}$. Growth in Petri dishes tilted by 2.5~mm, with \textbf{h,i} showing preferential expansion away from the tilt, indicating the crucial role of layer height in governing the dynamics. Scale bars, 10~mm (\textbf{a-e, h-i})}
    \label{figS:Height-variation}
\end{figure*}

\begin{figure*}[!]
    \centering
    \includegraphics[width=\textwidth]{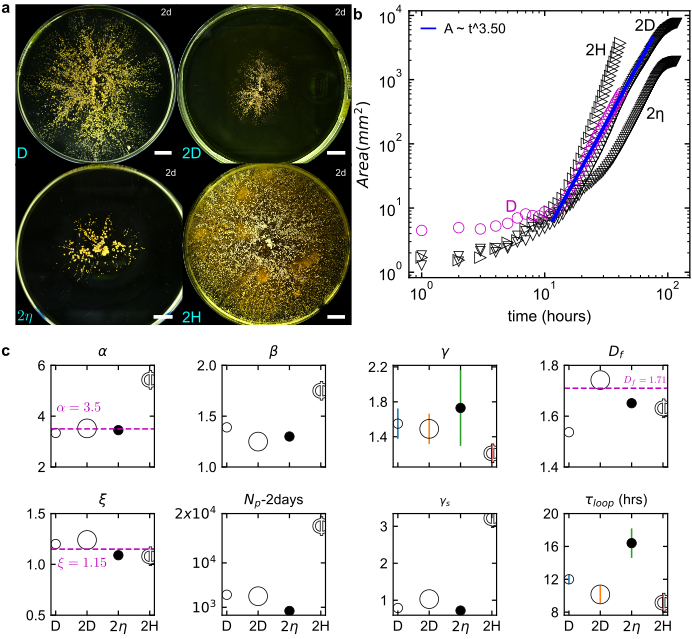}
    \caption{\textbf{Four variations in the viscous regime.} \textbf{a.} Firework patterns for variations in dish diameter ($D$, top row; same $H = 8$~mm, $\eta/\eta_w \sim 10^{4}$), viscosity ($\eta/\eta_w$, left column; same $H = 8$~mm, $D = 90$~mm), and layer height ($H$, right column; same $\eta/\eta_w \sim 10^{4}$, $D = 180$~mm) at 2~days (Supplementary Movie 3-6). \textbf{b.} Corresponding area growth curves for these conditions. \textbf{c.} Exponents and metrics associated with these variations: top-row--$\alpha$ (area growth), $\beta$ (crown diameter growth), $\gamma$ (single cluster growth), $D_f$--fractal dimension, bottom-row, $\xi$-cluster size distribution exponent, $N_{\text{particles}}$ at 2~days, $\gamma_{s}$-the seeding exponent, $\tau_{\text{loop}}$--the time it takes for a seed to complete a loop and land and stay at the bottom (Fig~\ref{figS:Controls-extras}\textbf{b}). Scale bars, 10~mm (\textbf{a}-left column), 20~mm (\textbf{a}-right column).}
    \label{figS:4cases-prt1}
\end{figure*}


\begin{figure*}[!]
    \centering
    \includegraphics[width=\textwidth]{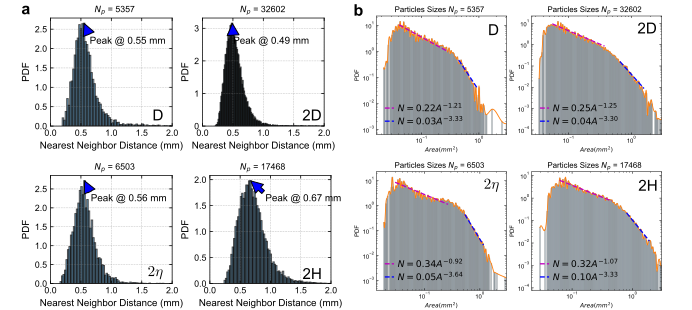}
    \caption{\textbf{Cluster statistics.} \textbf{a,b.} Nearest-neighbor distance and cluster size distributions for the four scenarios described in Fig.~\ref{fig:fig3}, after they nearly occupy the full petri-dish.}
    \label{figS:4-cases-extras2}
\end{figure*}

\begin{figure*}[!]
    \centering
    \includegraphics[width=\textwidth]{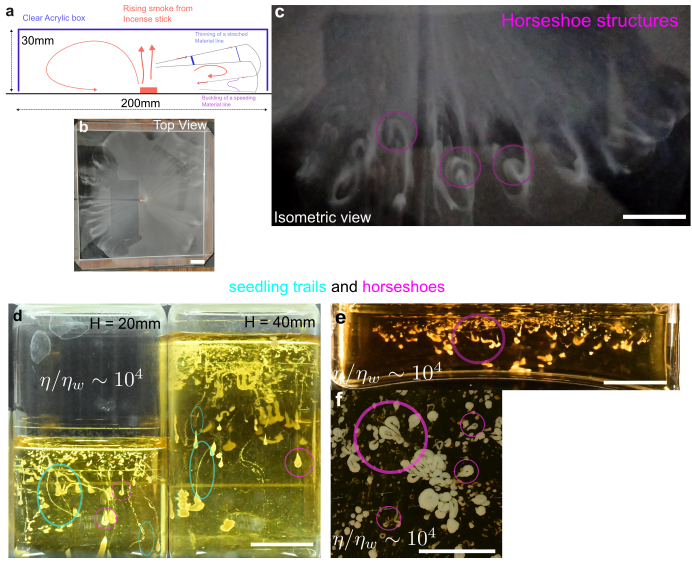}
        \caption{\textbf{Discerning the mechanism - Smoke, horseshoes, and seedlings.} \textbf{a.} Inverted-box setup showing a circulating smoke plume rising from an incense stick placed at the bottom (Supplementary Movie 8). \textbf{b,c.} Top and isometric views revealing horseshoe-like structures as the outward-moving smoke turns inward and descends near the container edges. \textbf{d,e.} Similar horseshoe patterns (pink circles) observed as yeast clusters loop back in their circulatory paths. Blue ovals mark fragmented seedlings formed as clusters move downward, initiating an auto-catalytic expansion (Supplementary Movie 2, 7). Scale bars, 20~mm (\textbf{b,c}), 10~mm (\textbf{d-f}).}
    \label{figS:Smoke}
\end{figure*}


\begin{figure*}[!]
    \centering
    \includegraphics[width=0.9\textwidth]{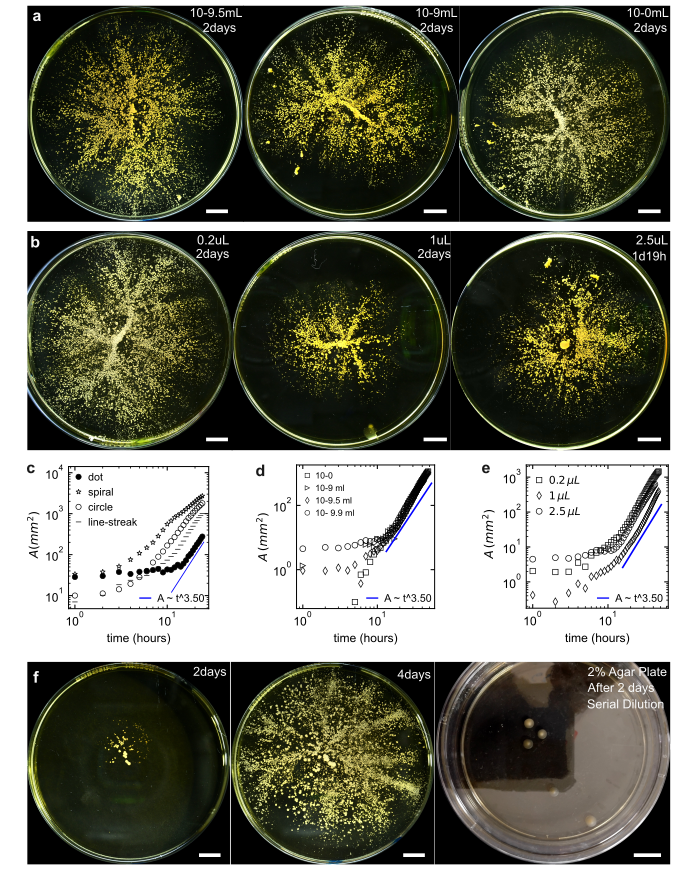}
    \caption{\textbf{Robustness to inoculation and initial conditions.} \textbf{a,b.} Firework patterns for varying inoculation cell densities (same 2.5~$\mu$L inoculation volume) and inoculation volumes (same $10$–$9.9$~mL) under the typical viscous condition ($\eta/\eta_w \sim 10^{4}$, $H = 8$~mm, $D = 90$~mm). \textbf{c.} Area growth curves for the varying initial conditions described in fig.~\ref{fig:fig4}\textbf{a}, with reference line $A \sim t^{3.5}$. \textbf{d,e.} Corresponding growth curves showing nearly identical area growth exponents, with reference line $A \sim t^{3.5}$. \textbf{f.} Firework formation (with a delay) observed even when starting from only a few cells. Scale bars, 10~mm (\textbf{a,b,f}).}
    \label{figS:Intial-Inoc Conditions}
\end{figure*}

\begin{figure*}[!]
    \centering
    \includegraphics[width=\textwidth]{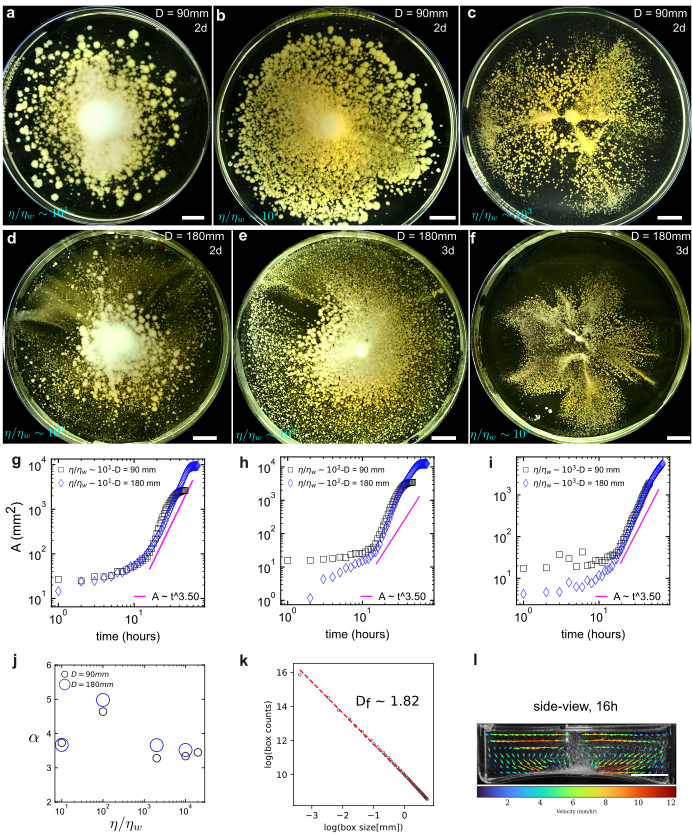}
    \caption{\textbf{Fireworks occur even at lower viscosities.} \textbf{a–c.} Firework patterns for varying viscosities ($\eta/\eta_w \sim 10^{1}$–$10^{3}$, $D = 90$~mm, $H = 8$~mm). \textbf{d–f.} Corresponding cases in a larger dish ($D = 180$~mm). \textbf{g–i.} Area growth curves for these conditions, with reference line $A \sim t^{3.5}$. \textbf{j.} Calculated area growth exponents. \textbf{k,l.} Fractal dimension ($D_f \sim 1.82$, indicative of more area fraction) and side-view PIV flow fields for the case showing a peak in the area-growth exponent ($\eta/\eta_w \sim 10^{2}$, Supplementary Movie 9). Scale bars, 10~mm (\textbf{a-c,l}), 20~mm (\textbf{d-f}).}
    \label{figS:eta-variation}
\end{figure*}

\begin{figure*}[!]
    \centering
    \includegraphics[width=\textwidth]{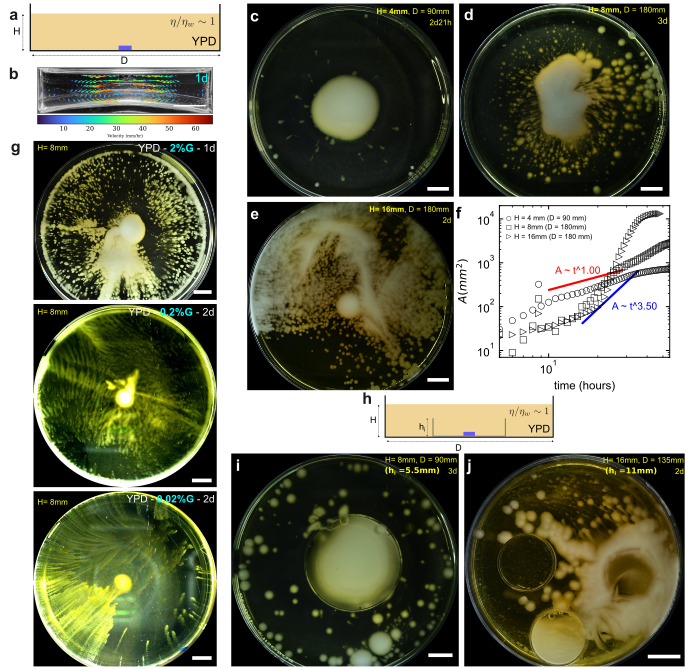}
    \caption{\textbf{Firework-like dispersal occurs even in water with yeast-growing-media, YPD ($\eta/\eta_w \sim 1$).} \textbf{a–b.} Similar to~fig.~\ref{fig:fig1}\textbf{b} but faster circulatory flows observed one day after inoculation in YPD. \textbf{c–e.} Slow to rapid expansions as the layer height varies from 4–16~mm, resembling behavior in the viscous regime~fig~\ref{fig:fig2}\textbf{f}. \textbf{f.} Corresponding area growth curves, with reference lines $A \sim t^{1}$ and $A \sim t^{3.5}$. \textbf{g.} Streaky-bicycle-spoke-like spreading observed as the sugar concentration is reduced from 2\% to 0.02\%. \textbf{h–j.} Inoculation within a smaller dish immersed in a larger one shows cells carried outward in circulatory loops, hopping over the rim of the inner dish—suggesting mechanisms similar to those at higher viscosities. Scale bars, 10~mm (\textbf{b,c,g,i}), 20~mm (\textbf{d,e,j}). }
    \label{figS:YPD-all}
\end{figure*}

\begin{figure*}[!]
    \centering
    \includegraphics[width=0.9\textwidth]{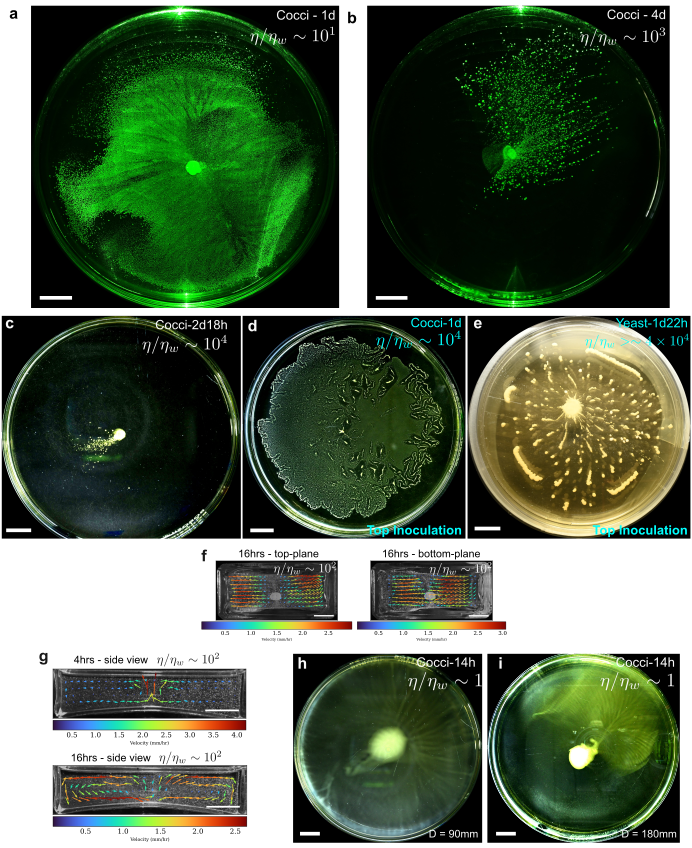}
    \caption{\textbf{Fireworks with tiny coccus bacteria ($\sim$1~$\mu$m, \textit{Staphylococcus aureus}).} \textbf{a–c.} Firework patterns at viscosities $\eta/\eta_w \sim 10^{1}$, $10^{3}$, and $10^{4}$, showing progressively slower spreads with increasing viscosity. \textbf{d–e.} Top inoculations of cocci and yeast showing rapid range expansions, similar to those reported in~\cite{atis2019microbial}, under viscous conditions ($\eta/\eta_w \sim 10^{4}$). \textbf{f–g.} Top- and bottom-plane PIV flow fields, along with side-view fields, indicating circulatory mechanisms analogous to those seen in yeast. \textbf{h–i.} Bicycle-spokes-like spreads of cocci inoculated in plain LB medium ($\eta/\eta_w \sim 1$). Layer height is same for all cases here, $H= \text{8~mm}$. Scale bars, 10~mm (\textbf{a-h}), 20~mm (\textbf{i}).}
    \label{figS:Cocci-all}
\end{figure*}


\begin{figure*}[!]
    \centering
    \includegraphics[width=\textwidth]{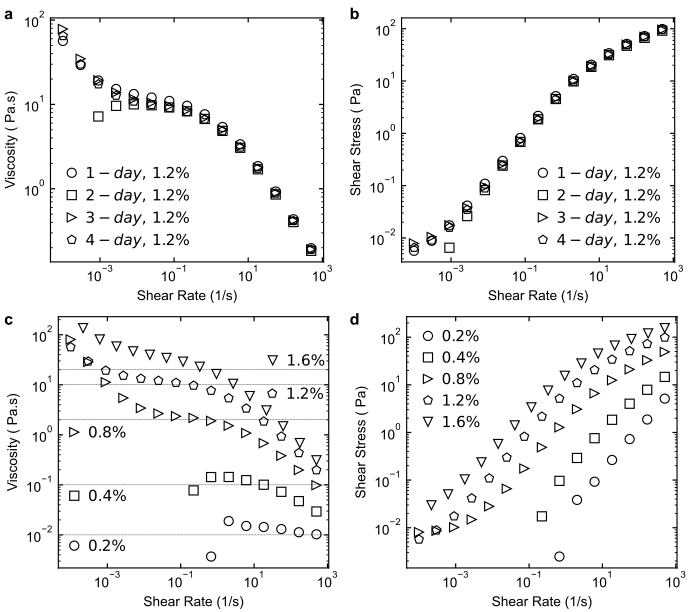}
    \caption{\textbf{Rheometry.} \textbf{a–b.} Viscosity  and shear stress variations with shear rate, measured over different days for the viscous substrate used in the typical case ($\eta/\eta_w \sim 10^{4}$), prepared by adding 1.2\% HEC to YPD (see Methods~\ref{methods:HEC-prep}). \textbf{c–d.} Viscosity (Pa·s) and shear stress (Pa) as a function of HEC concentration (0.2–1.6\%), spanning a range of $\eta/\eta_w \sim 10^{1}$–$2\times10^{4}$. }
    \label{figS:Rheometry}
\end{figure*}

\end{document}